# Hot carrier diffusion in graphene

Brian A. Ruzicka[1], Shuai Wang[2], Lalani K. Werake[1], Ben Weintrub[1], Kian Ping Loh[2], and Hui Zhao[1*]

[1]*Department of Physics and Astronomy, The University of Kansas, Lawrence, Kansas 66045, USA*
[2]*Department of Chemistry, National University of Singapore, 3 Science Drive 3, Singapore 117543*

We report an optical study of charge transport in graphene. Diffusion of hot carriers in epitaxial graphene and reduced graphene oxide samples are studied using an ultrafast pump-probe technique with a high spatial resolution. Spatiotemporal dynamics of hot carriers after a point-like excitation are monitored. Carrier diffusion coefficients of 11,000 and 5,500 squared centimeters per second are measured in epitaxial graphene and reduced graphene oxide samples, respectively, with a carrier temperature on the order of 3,600 K. The demonstrated optical techniques can be used for non-contact and non-invasive in-situ detection of transport properties of graphene.

## I. INTRODUCTION

Graphene, a single layer of carbon atoms, has superior charge transport[1–3], thermal transport[4], and mechanical properties.[5] These make graphene a very attractive candidate for applications like transistors,[1,6] solar cells,[7] electromechanical resonators,[8] ultracapacitors,[9] and composite materials.[10] Among these properties, charge transport is the most extensively studied one since it is the foundation of most applications.[1–3,6,11–20] Significant progress has been made in these studies, with demonstrations of ultrahigh mobilities at room temperature,[1–3] anomalous quantum Hall effects,[2,3,21,22] and a conductivity without charge carriers.[2]

One important aspect of charge transport in graphene is the role played by hot carriers. It has been shown that the mean-free path of carriers in graphene is several 100 nm even at room temperature.[1,11] Therefore, even in devices with a channel length as long as of 1 $\mu$m, injected carriers only undergo few or even no phonon scattering events during the transport. In transport measurements performed with an applied voltage on the order of a fraction of one volt, the carriers can have a temperature of several thousand kelvins.[20] To develop nanoscale devices of graphene, it is necessary to understand and control hot carrier dynamics.

Ultrafast laser techniques are standard tools to study hot carrier dynamics in semiconductors.[23] Over the past two years, significant progress has been made on using these techniques to study hot carriers in graphene.[24–34] In these experiments, hot carriers with well-defined energy distributions are quasi-instantaneously injected by interband absorption of an ultrafast laser pulse. The hot carrier dynamics is probed by time resolving the change of the transmission or reflection of a probe pulse in near infrared or terahertz spectral ranges. Carrier thermalization, energy relaxation, and recombination have been study in several types of graphene samples, including epitaxial graphene,[24–31] mechanically exfoliated graphene on Si/SiO$_2$ substrates,[32] graphene thin films grown by chemical vapor deposition,[30] reduced graphene oxide suspensions,[33] and reduced graphene oxide thin films.[34] However, due to the limited spatial resolution of these experiments, *transport* of hot carriers has not been studied.

In this Article, we report an high-spatial-resolution ultrafast pump-probe study of hot carrier transport in epitaxial graphene and reduced graphene oxide samples. In contrast to the previous transport studies by electric techniques, where mobilities of carriers under an externally applied electric field were measured,[1–3,6,11–19] we study diffusion of hot carriers driven by the density gradient, without applying an electric field. Carriers with a point-like spatial density profile is excited with an excess energy of more than 800 meV by a tightly focused ultrafast laser pulse through interband excitation. Expansion of the carrier profile is monitored by measuring differential transmission of a time-delayed and spatially scanned probe pulse. Carrier diffusion coefficients of 11,000 and 5,500 cm$^2$s$^{-1}$ are measured in epitaxial graphene and reduced graphene oxide samples, respectively, with a carrier temperature on the order of 3,600 K. The measured diffusion coefficients are compared with previously reported mobilities, by using the Einstein relation. It is quite encouraging that the hot carrier transport properties of the reduced graphene oxide sample is only a factor of two worse than the epitaxial graphene, since the fabrication of this type of graphene is low cost. The demonstrated optical techniques can be used for non-contact and non-invasive in-situ detection of transport properties of graphene.

## II. EXPERIMENTAL TECHNIQUES AND PROCEDURES

We choose two types of graphene samples for our study that are both of great technological relevance, epitaxial graphene and reduced graphene oxide. The former has great potentials to be used in semiconductor industry since it can be produced on large scales with a high degree of repeatability on a insulating substrate,[11] while the latter can be produced with low cost.[18] The epitaxial graphene samples are prepared on a Si-terminated 6H-SiC (0001) crystalline wafer surface by solid-state graphitization.[35,36] The reduced graphene oxide samples are fabricated by spin coating graphene oxide flakes on quartz substrates to form thin films, which are then transformed to graphene films by thermal reduction at 1,000°C. By using an atomic force microscope and a scanning tunneling microscope, we determine that the epitaxial samples have one or two layers of graphene, and the reduced oxide graphene samples contains about 50 layers.[34]

The experimental approach for the optical study of carrier transport is rather straightforward. Carriers are first excited with a pump laser pulse that is incident normal to the graphene layer, with a central wavelength of 750 nm and a pulse width of 100 fs, as illustrated in Fig. 1A. The pump pulse is obtained by frequency doubling the signal output of an optical parametric oscillator pumped by a Ti:Sapphire laser. From the pump photon energy, we estimate that the initial carrier temperature $T_e \approx 4,300$ K. The pump pulse is focused to a spot size of $w_0 = 1.6$ $\mu$m at full width at half maximum (FWHM) by using a microscope objective lens with a high numerical aperture. The spatial density profile of excited carriers is initially thin, but after a short time, the carriers diffuse out of the excitation spot, which results in a broadening of the profile (Fig. 1B). In this process, electrons (e) and holes (h) move as pairs due to the Coulomb attraction between them.

Such a classical diffusion process is described by the diffusion equation,

$$\frac{\partial n}{\partial t} = D \nabla^2 n - \frac{n}{\tau_r}, \qquad (1)$$

where $D$ is the diffusion coefficient and $\tau_r$ is the lifetime of the carriers. Since the pump laser spot has a Gaussian shape, the injected carriers have a Gaussian spatial profile within the graphene layer, $n(r,0) = Ne^{-4\ln(2)r^2/w_0^2}$, where $N$ is the peak carrier density. With this initial condition, the solution to the diffusion equations gives the density at a later time of[37]

$$n(r,t) = N \left( \frac{w_0^2}{w_n^2(t)} \right) e^{-4\ln(2)r^2/w^2(t) - t/\tau_r}, \qquad (2)$$

where

$$w^2(t) = w_0^2 + 16\ln(2)Dt. \qquad (3)$$

Clearly, the squared width of the profile expands linearly. By measuring $w$ as a function of time, the diffusion coefficient can be determined. Previously, such a technique has been widely used to study ambipolar diffusion in semiconductors.[38–44]

In our experiments, we monitor the diffusion process by using a time-delayed probe pulse with a central wavelength of 810 nm and a pulse width of 190 fs, obtained from the Ti:Sapphire laser that is used to pump the optical parametric oscillation. The probe pulse is focused by another microscope objective lens to a spot size of 1.2 $\mu$m. We verified that under our experimental conditions, the differential transmission of the probe pulse, $\Delta T/T_0 \equiv [T(n) - T_0]/T_0$, i.e. the normalized difference in transmission of the probe with $[T(n)]$ and without $(T_0)$ carriers, is proportional to $n$. A mechanical chopper is used to modulate the pump pulse for lock-in detection. In order to achieve a high signal-to-noise ratio, a balanced-detection scheme is used to suppress the common-mode laser intensity noise, with the reference pulse taken from the Ti:Sapphire laser.[45] Figure 1C shows the spatial profile of $\Delta T/T_0$ measured in the epitaxial graphene sample with the probe pulse arriving 0.08 ps after the pump pulse, as we scan the probe spot in the $xy$ plane by moving the microscope objective this focuses the probe pulse. With a peak energy fluence of the pump pulse of 170 $\mu$J cm$^{-2}$, the peak areal carrier density is estimated to be $10^{13}$ cm$^{-2}$. The Gaussian shape of the profile is consistent with the laser spots.

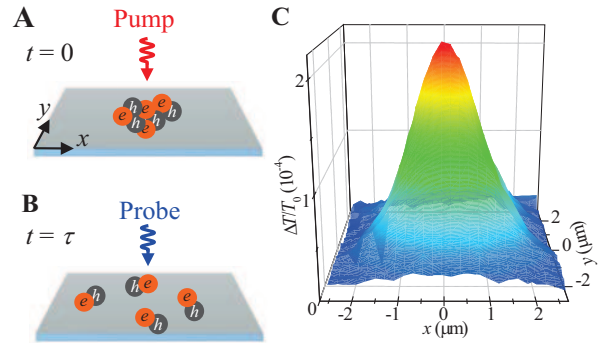

FIG. 1. A: A tightly focused pump laser pulse injects hot carriers in a graphene sample by exciting electrons (e) to the conduction band, leaving holes (h) in the valance band. (B) After a certain period of time, the injected carriers have diffused away from the excitation spot. The expanded carrier density profile is detected by a tightly focused probe pulse. (C) Spatial distribution of the differential transmission signal measured with the probe pulse arriving 0.08 ps after the pump pulse, on the epitaxial graphene sample at room temperature with an injected areal carrier density of $10^{13}$cm$^{-2}$ at the center of the excitation spot.

To quantitatively study the diffusion process, we measure the profiles along the $\hat{x}$ axis, where the signal is the strongest, for many probe delays, $\tau$. Figure 2A shows an example of the results of such scans. At each $x$, the $\Delta T/T_0$ decays rapidly with time. The red curve in Fig. 2B (left axis) shows a cross section of Fig. 2A at $x = 0$. Such a fast decay is consistent with the following picture of the carrier dynamics that has been established by previous time-integrated pump-probe experiments: After excitation, the carriers quickly reach a hot distribution via carrier-carrier scattering within a time scale shorter than 0.1 ps.[24–26,33,46] Then, the carriers cool through carrier-phonon scattering on a time scale on the order of 1 ps.[30] The decay of $\Delta T/T_0$ in Fig. 2B is mainly caused by carriers moving out of the detection window of the probe pulse in energy space. Since $\Delta T/T_0$ is proportional to the density of carriers at the probing energy, we can calculate how $T_e$ changes over time using the mea-sured $\Delta T/T_0$. The result is shown as the blue line in Fig. 2B (right axis).

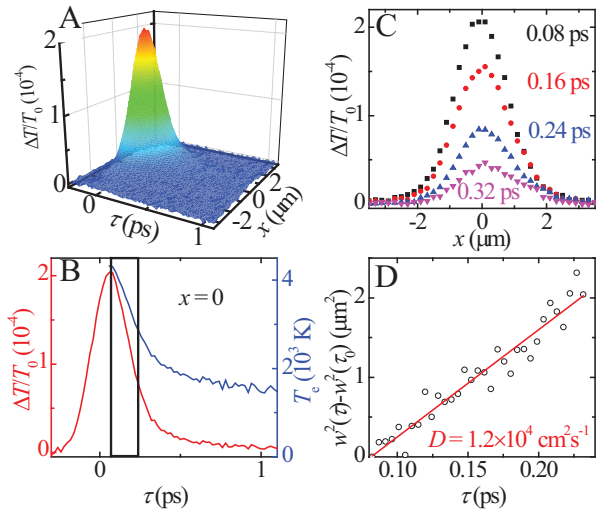

FIG. 2. A: The differential transmission signal as functions of the position of the probe spot with respect to the pump spot, $x$, and the delay time of the probe pulse with respect to the pump pulse, $\tau$. B: A cross section of Panel A with a fixed $x = 0$ (red line, left axis) and the deduced carrier temperature, $T_e$ (blue line, right axis). C: Cross sections of Panel A along $x$ for several values of $\tau$. D: The expansion of the squared width of the spatial profile. The solid line is a linear fit corresponding to a diffusion coefficient $D = 1.2 \times 10^4$ cm$^2$s$^{-1}$.

At all probe delays, the spatial profiles of $\Delta T/T_0$ along $x$ retain a Gaussian shape, with a few examples shown in Fig. 2C. By a Gaussian fit to the profile measured at each probe delay in the range from 0.08 to 0.24 ps (indicated as the box in Fig. 2B), we deduce the expansion in the squared width, $w^2(\tau) - w^2(\tau_0)$, where $w(\tau_0)$ is the width of the profile at probe delay $\tau_0 = 0.08$ ps, as shown in Fig. 2D. We choose this time range because the $\Delta T/T_0$ signal in this range is large enough for reliable measurements of $w$. From a linear fit (solid line), we deduce a diffusion coefficient of $D = 1.2 \times 10^4$ cm$^2$s$^{-1}$. We note that neither the decay of the signal nor the finite size of the probe spot influences the measurement of $D$: the former does not change the $w$, and the latter adds a constant, squared width of the probe spot, to the $w^2$ and thus does not change the slope. Since in the time range of the measurement, $T_e$ changes from 4,300 to 2,900 K, we have measured the $D$ of hot carriers with a temperature in that range.

It is worth noting that the measurement of $D$ is not influenced by the finite lifetime of carriers since the recombination of carriers only changes the height of the density profile, not the width.[47] Furthermore, the finite size of the probe spot does not influence the measurement neither. Since the probe spot size is comparable to the pump spot size, the profile measured by scanning the probe spot across the pump spot is actually convolutions of the probe spot and the actual carrier density profiles. However, since both the probe spot and the carrier density profiles are Gaussian, $w^2 = w_p^2 + w_N^2$, where $w_p$ and $w_N$ are the widths of the probe spot and the carrier density profile, respectively. Since $w_p^2$ exists in both sides of Eq. 3, the convolution doesn't influence the measurement of $D$.

We use this procedure to systematically investigate the carrier diffusion in both types of graphene samples. First, at sample temperatures of 300 K and 10 K, no change of the diffusion coefficient was observed when the carrier density is lowered by a factor of five. This indicates that the carrier-carrier scattering does not influence the diffusion process under our experimental conditions. The diffusion coefficient is also measured as a function of the sample temperature in the range of 10 - 300 K. The results are shown in Fig. 3 as the solid squares for the epitaxial sample and the solid circles for the reduced graphene oxide sample. At each temperature, multiple measurements were taken at different locations of the samples. The uncertainties on the deduced diffusion coefficient are caused by both the stability of the experimental setup and the inhomogeneity and reproducibility of the samples.

### III. RESULTS AND DISCUSSIONS

For each type of graphene samples, the diffusion coefficient does not show a significant dependence on the sample temperature. This is to be expected, however. Immediately after excitation the hot carriers emit a large amount of phonons during the fast energy relaxation process, causing the phonon distribution in the excitation spot to be extremely nonthermal. Since the local phonon distribution is determined by the hot carriers instead of the sample temperature, the environment that the carriers experience is independent of the temperature of the whole sample. By averaging the values measured at all the sample temperatures, we get the diffusion coefficients of $1.1 \times 10^4$cm$^2$ s$^{-1}$ for the epitaxial sample and $5.5 \times 10^3$cm$^2$ s$^{-1}$ for the reduced graphene oxide sample. It is quite encouraging that the hot carrier transport properties of the reduced graphene oxide sample is only a factor of two worse than the epitaxial graphene, since the fabrication of this type of graphene is low cost.[18]

In this optical study, equal numbers of electrons and holes are injected, and therefore the diffusion is ambipolar. The ambipolar diffusion coefficient is related to the unipolar diffusion coefficients of electrons and holes, and is dominated by the lower one.[48] Since the electrons and holes in graphene have the same energy dispersion, the diffusion coefficients for both are expected to be the same. Therefore, the diffusion coefficients we determined are also the unipolar diffusion coefficients of electrons and holes.

For a thermalized system, the diffusion coefficient is related to the mobility, $\mu$, by Einstein relation, $\mu =$

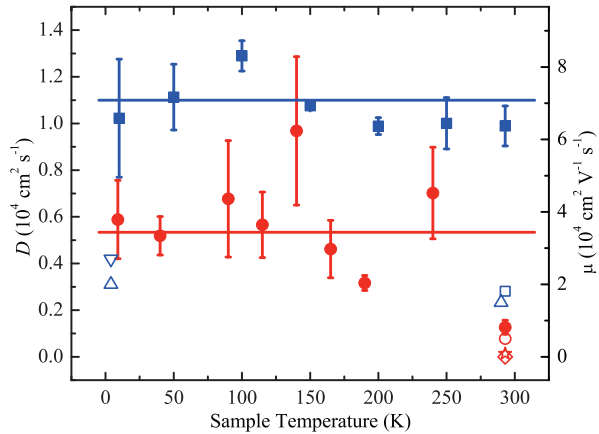

FIG. 3. Diffusion coefficient ($D$, left axis) as a function of the sample temperature for the epitaxial sample (solid squares) and the reduced graphene oxide sample (solid circles) measured by the procedure summarized in Fig. 2. The mobility ($\mu$, right axis) is deduced by using Einstein relation with a carrier temperature $T_e = 3600$ K. For comparison, previously reported values of mobilities in epitaxial graphene (the open square from Ref.[12], the open up-triangles from Ref.[13], and the open down-triangle from Ref.[11]) and reduced graphene oxide samples (the open circle from Ref.[17], the star from Ref.[19], and the diamond from Ref.[18]) are also plotted.

$2qD/k_B T_e$, where $q$ and $k_B$ are the amount of charge of each carrier and Boltzmann's constant. Due to the ultrafast thermalization process in graphene,[24–26,33,46] the carriers can be treated as a thermalized system. However, the diffusion occurs during the energy relaxation process. During the time range of the measurement, $T_e$ changes from 4,300 to 2,900 K (Fig. 2B, right axis). To estimate the mobility corresponding to the measured diffusion coefficients, we use an average temperature in this range of 3,600 K. The deduced mobilities are shown in Fig. 3 with the same symbols as the diffusion coefficient, but with the right axis. For the epitaxial graphene sample, we obtain a mobility of $7.0 \times 10^4$ cm$^2$ V$^{-1}$s$^{-1}$. Previous mobility measurements on epitaxial graphene have obtained rather different results.[6,11–16] However, the highest reported values are typically in the range of $1.5 \times 10^4$ to $2.7 \times 10^4$ cm$^2$ V$^{-1}$s$^{-1}$ (Fig. 3),[11–13] almost independent of sample temperature. We note that the mobility we deduced are for a carrier temperature on the order of 3,600 K, while the in the previous experiments the carrier temperatures are expected to be significantly lower. For the reduced graphene oxide sample, we deduced a mobility of $3.4 \times 10^4$ cm$^2$ V$^{-1}$s$^{-1}$. Fewer studies on mobility in this type of graphene have been reported; available data are plotted in Fig. 3.[17–19]

We have also attempted to study carrier dynamics in few-layer graphene samples prepared by chemical vapor deposition on metal substrates that were transferred to glass substrates. However, the differential transmission signal changes from positive at early time delays to negative at later time delays. Apparently, in addition to the phase-state filling, other mechanisms also contribute to the differential transmission signal. Since the differential transmission signal is not proportional to the carrier density, our procedure of measuring the diffusion coefficient is not valid for these samples.

## IV. SUMMARY

In summary, we have shown that an optical technique with high temporal and spatial resolution can be used to study transport of hot carriers in graphene and to directly measure the diffusion coefficients. With a carrier temperature on the order of 3,600 K, diffusion coefficients of $1.1 \times 10^4$ cm$^2$ s$^{-1}$ and $5.5 \times 10^3$ cm$^2$ s$^{-1}$ are determined in epitaxial graphene and reduced graphene oxide samples. It is quite encouraging that the hot carrier transport properties of the reduced graphene oxide sample is only a factor of two worse than the epitaxial graphene, since the fabrication of this type of graphene is low cost. Furthermore, the optical technique is non-contacting and non-invasive. It can be used for in situ detection of transport properties at different locations of a sample or for direct comparison of multiple samples. Since no electrode is needed, its potential influence on the transport measurement is excluded.

### ACKNOWLEDGEMENT

We acknowledge support from the US National Science Foundation under Awards No. DMR-0954486 and No. EPS-0903806, and matching support from the State of Kansas through Kansas Technology Enterprise Corporation. We thank the support of NRF-CRP "Graphene Related Materials and Devices" (Grant No. R-143-000-360-281). Acknowledgment is also made to the Donors of the American Chemical Society Petroleum Research Fund for support of this research.